\documentstyle[prc,aps,graphicx,epsf]{revtex}
\begin{document}
\title{
Hydrodynamic analysis of non-central \\
Pb+Pb collisions at 158\textit{A} GeV
}
\draft
\author{Tetsufumi Hirano, Keiichi Tsuda, and Kohei Kajimoto}
\address{Department of Physics, Waseda University, Tokyo 169-8555, Japan}
\date{\today}
\maketitle
\begin{abstract}
We analyze non-central heavy-ion collisions at the relativistic energy within a full (3+1) dimensional hydrodynamic model.
First, the initial parameters in the hydrodynamic model are chosen so that we reproduce the experimental data of both the rapidity distribution and the slope of transverse mass distribution in central Pb + Pb collisions at 158 $A$ GeV at the CERN SPS. We next study the validity of the ansatz of wounded nucleon scaling for the initial condition of the hydrodynamic model through analysis of the rapidity distribution of negative pions in non-central collisions.
Moreover, a simple Monte Carlo method is introduced to exactly calculate the particle distribution from resonance decays.
\end{abstract}

\section{INTRODUCTION}
The main goals in the physics of relativistic heavy-ion collisions are not only the discovery of a new state of deconfined nuclear matter, the quark-gluon plasma (QGP), but also the investigation of thermodynamical aspects of its new phase, i.e., the equation of state (EOS), the order of phase transition between the QGP phase and the hadron phase, or the critical temperature \cite{QM99}. 
The hydrodynamic model is appropriate for this purpose since thermodynamical variables such as temperature, energy density, and pressure are basic ingredients of the model. 
Hydrodynamics is directly connected with the EOS of nuclear matter, thereby we describe the phase transition phenomena in a straightforward manner in terms of hydrodynamics; on the other hand, it is somewhat difficult to discuss the confinement-deconfinement phase transition by microscopic kinetic theories.
Therefore, if hydrodynamics is applicable to the description of space-time evolution of nuclear matter, it has some advantages over microscopic kinetic theories when we investigate the thermodynamics of the QGP. 

In addition, the hydrodynamic model enables us to analyze the collective flow of nuclear matter produced in heavy-ion collisions. 
In the central Pb + Pb collisions at 158 $A$ GeV at the CERN Super Proton Synchrotron (SPS), the inverse slope parameters of the transverse mass spectra for the non-multistrange hadrons $\pi$, $K$, $p$, and $d$ are parametrized by the two common values, the freeze-out temperature and the radial velocity \cite{NA44}.
This implies that these particles constitute the radial flow, and that the local thermalization among those particles is achieved in \textit{central} collisions.
The experimental results suggest the hydrodynamic description for the expansion of nuclear matter seems to be suitable in central collisions at the SPS energy.
On the other hand, it is an open question whether the hydrodynamic picture is valid even in \textit{non-central} collisions.

As well as central heavy-ion collisions, non-central events contain many attractive phenomena; these are important in understanding the phase transition between the QGP phase and the hadron phase. 
From the hydrodynamic point of view, there are two interesting topics in non-central collisions.
One is the elliptic flow, which is one of the anisotropic collective flows on the transverse plane.
Fourier analysis of azimuthal distribution for measured particles is often used to discuss the anisotropic transverse flow quantitatively \cite{POS98}.
The first and the second Fourier coefficients stand for the directed and the elliptic flow respectively.
Ollitrault showed that the elliptic anisotropy is largely affected by the EOS of nuclear matter \cite{OLLI92}.
Sorge also showed that elliptic flow is highly sensitive to pressure at the early stage of heavy-ion collisions \cite{SORGE}.
Heiselberg and Levy discussed that the centrality dependence of elliptic flow may have an unusual structure when the matter at the early stage crosses the phase transition region as varying the impact parameter \cite{HEISEL}.
Voloshin and Poskanzer showed that the centrality dependence of elliptic flow is a good indicator for the thermalization of nuclear matter \cite{VOLO00}.
Experimentally, the elliptic flow was observed at CERN SPS \cite{NA49ELLI,WA98}.
The NA49 Collaboration found that the second Fourier coefficient of azimuthal distribution for charged pions is about 4 \% around the midrapidity and the low transverse momentum \cite{NA49ELLI}.
The WA98 Collaboration also obtained a similar result for positive pions.
In addition to this result, they observed a remarkable phenomenon for positive kaons; only positive kaons are emitted out of the reaction plane at the SPS energy \cite{WA98}.
The other topic in non-central collisions is the ``nutcrack" phenomenon advocated by Teaney and Shuryak \cite{NUT}.
According to their results, the unusual structure of nuclear matter can appear  if the QCD phase transition occurs in non-central collisions at Relativistic Heavy-Ion Collider (RHIC) energies.
Along the lines of these works, Kolb \textit{et al.}~studied these two topics by using hydrodynamic simulations \cite{KOLB}.
The theoretical studies mentioned above are, however, all based on the (2+1) dimensional hydrodynamic model \cite{OLLI92,NUT,KOLB} or on microscopic kinetic theories \cite{SORGE,HEISEL,VOLO00}.
Therefore, full (3+1) dimensional hydrodynamic simulations are indispensable
for a comprehensive understanding of the expansion stage of heavy-ion collisions and the phase transition of nuclear matter in the full phase space.
Here we study non-central heavy-ion collisions by using a fully three-dimensional relativistic hydrodynamic model. 
In this paper, we compare the numerical results of the rapidity distribution with the experimental data in non-central collisions at the SPS energy, before discussing each topic mentioned above. 
This work is the first analysis by means of a full (3+1) dimensional hydrodynamic model.
From a hydrodynamic standpoint, it is very important to choose the initial condition to reproduce the single particle spectra of hadrons since the predictive power of hydrodynamics is rather limited.
The hydrodynamic analyses of flow, dilepton spectra, HBT, and so on, are really meaningful only after tuning the hydrodynamic model.

The paper is organized as follows: In Sec.~\ref{RHM}, we briefly review the relativistic hydrodynamic model.
The relativistic hydrodynamic equations with a relevant EOS are introduced in this section.
We also discuss the initial condition of the hydrodynamic model to describe the non-central heavy-ion collisions.
In Sec.~\ref{PD}, we show how to calculate the particle distribution within the hydrodynamic model.
In particular, we focus on the contribution from resonance decays to the particle spectra.
In Sec.~\ref{HS}, we show the numerical results of hadron spectra in central and non-central collisions.
We examine in this section whether the ansatz of wounded nucleon scaling, which is widely used in the two-dimensional hydrodynamic model \cite{OLLI92,NUT,KOLB}, is really valid for the initial condition. 
Our results are summarized in Sec.~\ref{SD}.
In the Appendix, we briefly show how to obtain the contribution from resonance decays by using a Monte Carlo method.


\section{RELATIVISTIC HYDRODYNAMIC MODEL}\label{RHM}

\subsection{Relativistic hydrodynamic equation}
Assuming local thermal equilibrium for nuclear matter produced in heavy-ion collisions, we describe its space-time evolution by a relativistic hydrodynamic model.
Relativistic hydrodynamic equations for a perfect fluid represent energy and momentum conservations
\begin{eqnarray}
\label{EP}
\partial_{\mu}T^{\mu \nu}(x) &=& 0, \\
T^{\mu \nu}(x) &=& [E(x)+P(x)]u^\mu(x) u^\nu(x)-P(x) g^{\mu \nu},
\end{eqnarray}
and baryon density conservation
\begin{eqnarray}
\label{N}
\partial_{\mu}n_{B}^\mu(x) &=& 0, \\
n_{B}^\mu(x) &=& n_{B}(x)u^\mu(x), 
\end{eqnarray}
\noindent
where $E$, $P$, $n_{B}$, and $u^\mu = (\gamma, \gamma {\bf v})$ are, respectively, energy density, pressure, baryon density, and local four velocity.

\begin{figure}
\begin{center}
\includegraphics[width=12cm]{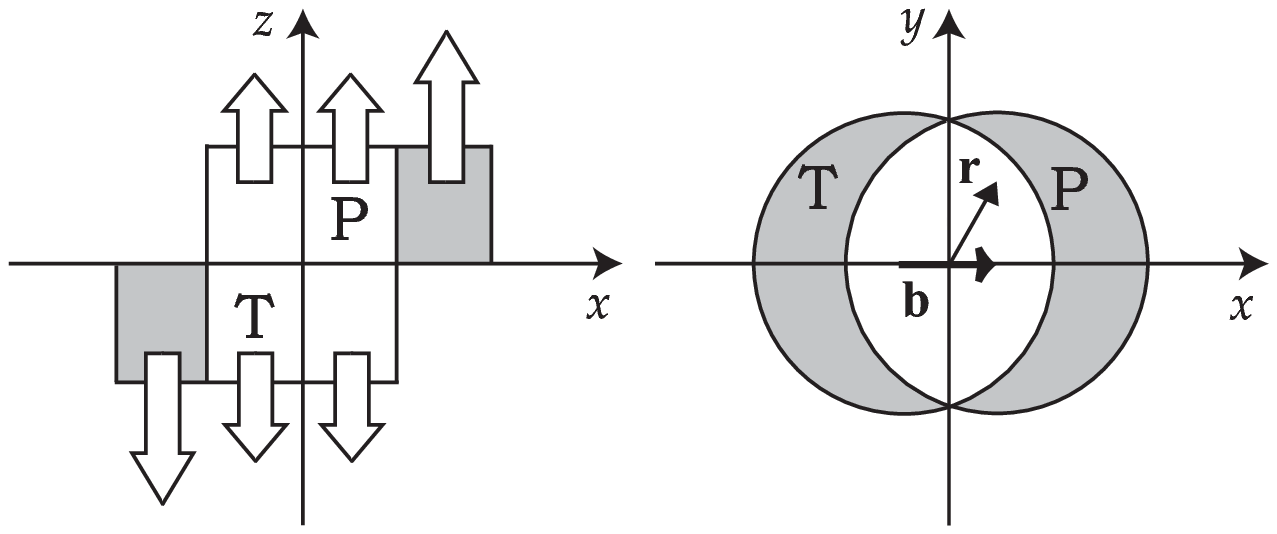}
 \caption{Schematic view of a non-central heavy-ion collision at the initial time $t_0$. The left (right) figure represents the reaction (transverse) plane. ${\bf b}$ is an impact parameter vector. The gray areas represent the spectators which we neglect in the present paper.}
 \label{FIG1}
\end{center}
 \end{figure}

When one analyzes central heavy-ion collisions, one can impose the cylindrical symmetry along the collision axis on the hydrodynamic equations.
The numerical simulations of the hydrodynamic model, with cylindrical symmetry, are relatively easy to perform due to the reduction of spatial dimension.
Hence some groups \cite{WASEDA,HUNG,SOLL97,ORNIK,DUMITRU} have studied the hadron spectra in central collisions at the SPS energies by their own hydrodynamic models without using Bjorken's scaling solution \cite{BJORKEN} which helps to simplify the solution.
On the other hand, in order to analyze non-central collisions, and moreover, to obtain the rapidity dependence of physical quantities, we numerically solve the hydrodynamic equations without assuming cylindrical symmetry along the collision axis or Bjorken's scaling solution \cite{BJORKEN}. 
We adopt Cartesian coordinates and rewrite the hydrodynamic equations (\ref{EP}) and (\ref{N}) as follows:
\begin{eqnarray}
\label{HE}
\partial_t \left(\begin{array}{c}
             U_1 \\
             U_2 \\
             U_3 \\
             U_4 \\
             U_5 \\ 
            \end{array} \right)
& + & \nabla \left(\begin{array}{c}
             U_1 \\
             U_2 \\
             U_3 \\
             U_4 \\
             U_5 \\ 
            \end{array} \right) {\bf v}
+  \left(\begin{array}{c}
             \partial_x P \\
             \partial_y P \\
             \partial_z P \\
             \nabla P {\bf v}\\
             0\\ 
            \end{array} \right) = 0,
\end{eqnarray}
where
\begin{eqnarray}
\label{U}
\left(\begin{array}{c}
             U_1 \\
             U_2 \\
             U_3 \\
             U_4 \\
             U_5 \\ 
            \end{array} \right) & = &
\left(\begin{array}{c}
             \gamma^2 (E+P)v_x \\
             \gamma^2 (E+P)v_y \\
             \gamma^2 (E+P)v_z \\
             \gamma^2 (E+P)-P \\
             \gamma n_B \\ 
            \end{array} \right).
\end{eqnarray}
As shown in Fig.~\ref{FIG1},
we can define the direction of $x$-axis as the impact parameter vector and that of $z$-axis as the collision axis.
There are two spatial symmetries with respect to the transformations $y \rightarrow -y$ and/or $(x, z) \rightarrow (-x, -z)$ for the system even in non-central collisions.\footnote{The latter symmetry exists only in symmetric collisions such as Pb + Pb or Au + Au.}
Because of the above symmetries, we numerically solve Eq.~(\ref{HE}) only in the region $-x_{\mathrm{max}} \le x \le x_{\mathrm{max}}$, $0 \le y \le y_{\mathrm{max}}$, and $ 0 \le z \le z_{\mathrm{max}}$.
We take $x_{\mathrm{max}}$ and $y_{\mathrm{max}}$ to be sufficiently larger than the nuclear diameter.
The longitudinal size in a numerical simulation depends on the lifetime of the fluid: $z_{\mathrm{max}} > t_f$. 
These spatial symmetries considerably reduce the computational time of hydrodynamic simulations.
If needed, it is easy to obtain the solutions in the other region $y<0$ and/or $z<0$ by the above transformations.

The algorithm we used to solve the hydrodynamic equations is the Piecewise Parabolic Method (PPM) \cite{PPM}.
This method is a kind of Godunov scheme.
The PPM is known as a very robust scheme for solving a non-relativistic gas equation with a shock wave.
We have extended the PPM scheme of Eulerian hydrodynamics to relativistic hydrodynamic equations. 
This enables us to describe precisely the space-time evolution of nuclear matter with phase transition.
It should be noted that the PPM is a higher order extension of the piecewise \textit{linear} method such as the rHLLE algorithm \cite{RHLLE}.

\subsection{Equation of state}
\label{EOS}
In order to solve the hydrodynamic equations, we need to specify the equation
of state, especially, pressure as a function of energy density and baryon density $P(E, n_B)$.
Since the hydrodynamic flow is accelerated by the pressure gradient, the space-time evolution of fluid is governed by the model EOS.
If the QGP phase appears in nuclear collisions, the quark matter expands, cools down, and crosses the softest point \cite{SOFT}.
The acceleration of a fluid element is considerably suppressed near the softest point since the ratio of the pressure to the energy density takes its minimum as a function of the energy density at this point.
Consequently, the hydrodynamic flow at freeze-out is expected to be reduced.
The softest point usually lies on the boundary between the QGP phase and the hadron phase on the $T$-$\mu$ plane. 
Therefore it is expected that the hadronic observables concerning the hydrodynamic flow contain information about the EOS of nuclear matter near the phase transition region \cite{OLLI92,SORGE,SOFT,RISCHKE}.

\begin{figure}
\begin{center}
\includegraphics[width=12cm]{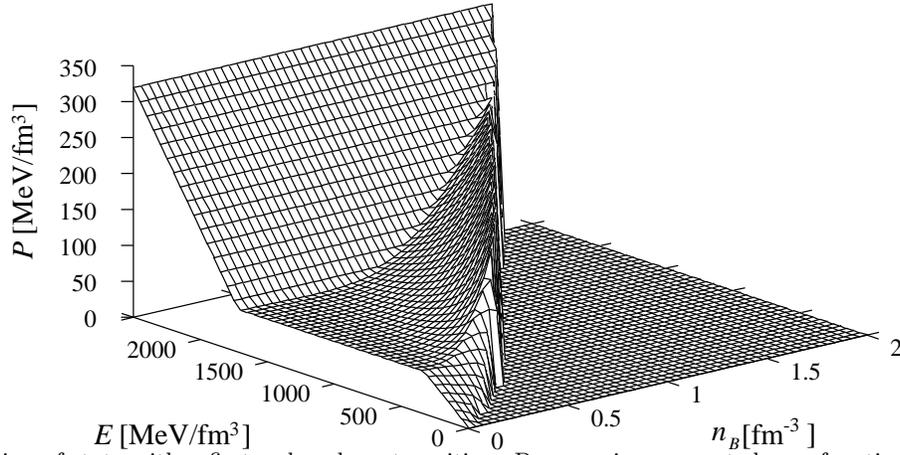}
 \caption{The equation of state with a first order phase transition. Pressure is represented as a function of energy density $E$ and baryon density $n_B$.}
 \label{FIG2}
\end{center}
 \end{figure}

We numerically solve Eq.~(\ref{HE}) under a relevant EOS and obtain $E$, $n_B$, and $v_i$ at each space-time point from the numerical solutions $U_i$ by using $P(E, n_B)$ and Eq.~(\ref{U}).
For an EOS, we adopt the same model as the one in Ref.~\cite{NONAKA}.
The model shows strong first order phase transition between the QGP phase and the hadron phase.
We assume, for simplicity, the QGP phase consists of massless and free $u$, $d$, $s$ quarks, and gluons.
On the other hand, we adopt the resonance gas model for the hadron phase.
The hadron phase is assumed to be composed of mesons and non-strange baryons up to the mass 2 GeV \cite{PDG}.
It is well known that pressure in the resonance gas model overcomes that of the QGP phase in high baryon density and low temperature region \cite{CLEYMANS}.
To avoid this problem, we also adopt an excluded volume correction \cite{XV} to suppress the pressure in the high baryon density region.
We construct a first order phase transition model by the usual technique of Gibbs' phase equilibrium condition, $T_{\mathrm{QGP}}=T_{\mathrm{hadron}}$ (thermal equilibrium), $\mu_{\mathrm{QGP}}=\mu_{\mathrm{hadron}}$ (chemical equilibrium), and $P_{\mathrm{QGP}}=P_{\mathrm{hadron}}$ (kinetic equilibrium).
A bag constant $B^{1/4} = 233$ MeV is chosen so that the phase transition occurs at the critical temperature 160 MeV at zero baryon density. 
The resultant pressure as a function of energy density and baryon density is represented in Fig.~\ref{FIG2}. Due to the Fermi energy of baryons, there exists a minimum energy density $E_{\mathrm{min}}(n_B)$ at a fixed baryon density.
For an illustrative purpose, we redefine the pressure as zero in the low energy density and high baryon density region in Fig.~\ref{FIG2}.
The latent heat $\Delta E(n_B=0)$ in our first order phase transition model is 1.4 GeV/fm$^3$.

The nuclear matter produced in Pb + Pb collisions may be isospin asymmetric since the lead nucleus is composed of 126 neutrons and 82 protons.
As well as other works based on the hydrodynamic model,
we assume that our model EOS can be used for the isospin asymmetric matter produced in Pb + Pb collisions.

\subsection{Initial conditions}\label{INI}
It is difficult to describe initial hard $NN$ collisions and pre-thermalization of the excited matter by using a one-fluid model, since hydrodynamics for a perfect fluid is applicable only to equilibrated matter.\footnote{It should be noted that Brachmann \textit{et al.}~studied, within the \textit{three}-fluids model, the whole of the stage of heavy-ion collisions including the compression of nuclear matter \cite{3FLUID}. By using this model, they discussed entropy production at the compression stage.}
Hence we apply the hydrodynamic model after the system is supposed to be thermalized.
We can choose the initial time $t_0$ when the incident Lorentz contracted nucleus finishes penetrating the target nucleus in the center of mass system.
In the Pb+Pb 158 $A$ GeV collision, we obtain $t_0 = z_0/v_0 = 2 R /(v_0 \gamma) \sim 1.44$ fm. Here $v_0$ is the velocity of incident nucleus in the center of mass system, $R$ is a radius of lead nucleus, and $z_0 = 2R/\gamma$ is a Lorentz contracted nuclear diameter. 

At the initial time $t_0$, we suppose the initial energy density and baryon density are factorized into the longitudinal and transverse parts as follows:
\begin{eqnarray}
E(t_0,x,y,z;b) & = & E(t_0,z) W(x,y;b),\\
n_B(t_0,x,y,z;b) & = & n_B(t_0,z) W(x,y;b).
\end{eqnarray}
Here $b$ is a impact parameter and $W$ is a transverse profile function.
We can parametrize the $z$ dependence of the initial energy density as Bjorken's solution \cite{BJORKEN}:
\begin{eqnarray}
\label{E}
E(t_0, z) = E_0 \left(\frac{\sqrt{t_0^2-z^2}}{t_0}\right)^{-\frac{4}{3}}\theta(\tilde{z}_0-\mid z \mid),
\end{eqnarray}
where we assume almost all the matter is in the QGP phase at $t_0$ and its phase obeys the ideal gas EOS, $P(E, n_B) = E/3$.
Since Bjorken's solution diverges on the light cone by definition, we introduce the additional factor $\theta(\tilde{z}_0-\mid \!z \!\mid)$ so that the system has finite energy.
An adjustable parameter $\tilde{z}_0 (< z_0)$ is to be chosen so that we reproduce the rapidity distribution of negative hadrons.
We emphasize that we have employed Bjorken's solution as an initial condition.
This is in contrast to Refs. \cite{OLLI92,NUT,KOLB}, in which Bjorken's boost invariant solution was always assumed to be valid during expansion of nuclear matter.
In Bjorken's picture, the resultant physical variables do not depend on the rapidity $Y$ anymore \cite{BJORKEN}. 
We also suppose that the baryon density as a function of $z$ is represented by
\begin{eqnarray}
\label{NB}
n_B(t_0, z) = n_{B0} \left(\frac{\sqrt{t_0^2-z^2}}{t_0}\right)^{-\alpha}\theta(\tilde{z}_0-\mid z \mid).
\end{eqnarray}
For the scaling solution, one takes $\alpha = 1$; while we take $\alpha$ to be larger than unity in order to obtain the two peaks in the rapidity distribution of net proton, which are seen in the experimental data at the SPS energy \cite{NA49RAP}. 
Initial longitudinal flow is parametrized by \cite{SOLL97}
\begin{eqnarray}
\label{VZ}
v_z(t_0, z) = \tanh \left(\frac{z}{t_0} \right)\theta(\tilde{z}_0-\mid z \mid).
\end{eqnarray}
This parametrization for the longitudinal velocity is different from Bjorken's solution: $v_{\mathrm{Bj}}(z) = z/t_0$.
As for the initial longitudinal velocity, this parametrization, rather than Bjorken's solution, enables us to reproduce the rapidity distribution well.
We also assume transverse flow vanishes at the initial time: $v_x(t_0) = v_y(t_0) = 0$. This means that transverse flow is produced and accelerated only by the pressure gradient perpendicular to the collision axis.

The transverse density profile at $t_0$ is supposed to be proportional to the number of wounded nucleons $N_W$ in a collision \cite{OLLI92,NUT,KOLB}: 
\begin{eqnarray}
\label{TR}
&&\frac{d N_W}{d^2 {\bf r}}({\bf r}; {\bf b}) 
 = T_a \left({\bf r}+\frac{{\bf b}}{2}\right)
\left\{1-\left[1-\frac{\sigma_{\mathrm{in}}
T_b({\bf r}-\frac{{\bf b}}{2})}{A}\right]^A\right\} 
 + T_b\left({\bf r}-\frac{{\bf b}}{2}\right)
\left\{1-\left[1-\frac{\sigma_{\mathrm{in}}
T_a({\bf r}+\frac{{\bf b}}{2})}{A}\right]^A\right\},
\end{eqnarray}
\noindent
where $T_i$ is the thickness function for nucleus $i$
\begin{eqnarray}
T_i({\bf r})=\int dz \rho\left(\sqrt{{\bf r}^2+z^2}\right),
\end{eqnarray}
and ${\bf r}$ is a vector on the transverse plane (see also Fig.~\ref{FIG1}). 
For nuclear density distribution, we use a Woods-Saxon distribution
\begin{eqnarray}
\rho(r) = \frac{\rho_0}{\exp[(r-R)/\delta]+1},
\end{eqnarray}
where $\delta = 0.5$ fm, $R =1.12 \times A^{1/3}$ fm, and $\rho_0 = 0.159$ fm$^{-3}$. $\sigma_{\mathrm{in}}$(= 30 mb) represents the $NN$ inelastic cross section at $\sqrt{s_{NN}} \sim 20$ GeV.
We define a transverse profile function $W(x, y; b)$ which is normalized to have unity at the origin in the case $b=0$ fm, i.e., $W(x, y; b) = \frac{d N_W}{d^2 {\bf r}}({\bf r}; {\bf b})/\frac{d N_W}{d^2 {\bf r}}({\bf 0}; {\bf 0})$.
By shifting $b$ in $W(x, y; b)$ and without changing other parameters in Eqs.~(\ref{E})-(\ref{VZ}), we easily obtain the initial conditions for non-central collisions specified by the impact parameter $b$.
It should be noted that the spatial distribution of initial collisions on the transverse plane is discussed in Ref.~\cite{JACOBS} and the procedure to obtain the density of interactions by shifting $b$ is called wounded nucleon scaling. In this paper, we use the same form for the parametrization of transverse profile in the initial condition.
The initial baryon density at origin $n_{B0}$ is chosen so that the number of wounded nucleons equals the number of baryons in the fluid:
\begin{eqnarray}
\label{NB0}
\int d^2 {\bf r}\frac{d N_W}{d^2 {\bf r}}({\bf r}; {\bf 0}) = \int dxdydz\ n_B(t_0, x, y, z; 0).
\end{eqnarray}
Taking into account the spectators is beyond the scope of the present paper. Hence we assume that the spectators leave from the excited matter instantaneously and that its effects on the space-time evolution of hot matter or on the particle distribution are neglected.

\section{PARTICLE DISTRIBUTIONS}\label{PD}
\subsection{Particle directly from freeze-out hyper-surface}
We assume that the hydrodynamic picture is valid above freeze-out energy density $E_{\mathrm{f}}$ and that
the particles do not interact with each other below $E_{\mathrm{f}}$.\footnote{The decay processes of resonance particles, however, exist. See the next subsection.}
Thus we can define the freeze-out hyper-surface $\Sigma$ and its elements $d \sigma^\mu$.
The numerical results of hydrodynamic simulation, i.e., temperature $T(x)$, chemical potential $\mu(x)$, the local four velocity $u^\mu(x)$, and the volume element $d \sigma_\mu(x)$ on the freeze-out hyper-surface $\Sigma$, give us the momentum distribution through the Cooper-Frye formula\footnote{It is well known that this formula has a problem although it has been widely used to obtain the momentum distribution.
When one applies the formula to the space-like freeze-out hyper-surface, it counts the particles reabsorbed by the fluid. Nevertheless we use this formula for simplicity.}\cite{CF}
\begin{eqnarray}
E\frac{dN}{d^3 p} = \frac{d}{(2 \pi)^3}\int_\Sigma \frac{p^\mu d \sigma_\mu(x)}{\exp\left\{\left[p^\nu u_\nu(x)-\mu(x)\right]/T(x)\right\} \mp 1}. 
\end{eqnarray}
Here $d$ is the degree of freedom for the particles considered in the spectrum.
By using this formula, we obtain the rapidity distribution
\begin{eqnarray}
\frac{dN}{dY} = \int d\phi p_t dp_t E\frac{d N}{d^3 p},
\end{eqnarray}
and the transverse mass distribution
\begin{eqnarray}
\frac{1}{m_t}\frac{dN}{dm_t dY} = \int d\phi E\frac{d N}{d^3 p}.
\end{eqnarray}
We apply this formula for the particles directly emitted from the freeze-out hyper-surface.

\subsection{Particles feeding from resonance decays}

In addition to the particles directly emitted from the freeze-out hyper-surface, resonance decays also contribute to the observed spectrum.
Assuming that resonance particles are emitted from freeze-out hyper-surface and that these particles decay in the vacuum, 
the multiplicity of pions through two-body decay processes
is given by \cite{SOLL91,HIRANO}
\begin{eqnarray}
\label{MULTI}
&&N_{R \rightarrow \pi X}  = 
\int \frac{J(p_l,\phi;{\bf V}_{R})dp_l d\phi}{4\pi p^*}
B_{R \rightarrow \pi X} \frac{d^3 {\bf p}_{R}}{E_{R}}
 \int ds W_{R}(s) \nonumber \\
&&\times\frac{d_{R}}{(2 \pi)^3}\int_\Sigma 
\frac{p_{R}^\mu d\sigma_\mu(x)}{\exp\left\{[p_{R}^\nu u_\nu(x) - \mu(x)]/T(x)\right\} \mp 1},
\end{eqnarray}
where $p_l$ and $\phi$ are, respectively, the longitudinal momentum and the azimuthal angle of pions. 
$p_R^\nu$ and $d_R$ are, respectively, 
the resonance four momentum in the reference frame and the degeneracy.
For boson (fermion) resonances, we take a $-(+)$ sign. 
$B_{R \rightarrow \pi X}$ is the branching ratio of
the decay process and $W_{R}$
is the Breit-Wigner type function, which takes account 
of the finiteness of the resonance width. For $W_{R}$, we adopt the
form used in Ref.~\cite{SOLL91}.
The Jacobian of the Lorentz transformation from the resonance
rest frame to an arbitrary reference frame $J(p_l, \phi; {\bf V}_{R})$ is defined by
$dp_l^* d\phi^* = J(p_l,\phi;{\bf V}_R)dp_l d\phi$,
where the quantities with (without) $*$ are the ones in the resonance
rest (reference) frame.
In the resonance rest frame, the momentum of pions is given by
\begin{eqnarray}
\label{PSTAR}
p^* = \frac{1}{2 m_R}\sqrt{(m_R + m_\pi)^2-m_X^2}\sqrt{(m_R - m_\pi)^2-m_X^2}.
\end{eqnarray}
We apply Eq.~(\ref{MULTI}) to the multiplicity of kaons or protons which emerge from two-body resonance decays. 
We note that the $\phi$ dependence of $J(p_l, \phi; {\bf V}_R)$ plays an important role in understanding the observed elliptic flow of pions \cite{HIRANO}.

When the resonance particle decays into three particles, i.e., $\omega \rightarrow 3\pi$, we write down the multiplicity of pions as follows \cite{SOLL91}:
\begin{eqnarray}
N_{R \rightarrow \pi X_1 X_2}& = & \int J(p_l,\phi;{\bf V}_{R})dp_ld\phi B_{R \rightarrow \pi X_1 X_2}
\int\limits_{\left(\sum\limits_{i=1,2} m_i \right)^2}^{\left(m_{R}-m_\pi \right)^2} dW^2
\frac{m_{R}\sqrt{W^2-(m_1+m_2)^2}\sqrt{W^2-(m_1-m_2)^2}}{2\pi Q(m_{R},m_i) W^2} \nonumber \\
 & \times &  \frac{d^3 {\bf p}_{R}}{E_{R}} \frac{d_{R}}{(2 \pi)^3}\int\frac{p_{R}^\mu d\sigma_\mu(x)}{\exp[p_{R}^\nu u_\nu(x)/T(x)]-1},
\end{eqnarray}
\noindent
where
\begin{eqnarray}
Q(m_R, m_i)  =  \int\limits_{(m_1+m_2)^2}^{(m_{R}-m_\pi)^2}\frac{dx}{x} 
\sqrt{(m_R+m_\pi)^2-x}\sqrt{(m_R-m_\pi)^2-x}
\sqrt{x-(m_1+m_2)^2}\sqrt{x-(m_1-m_2)^2}.
\end{eqnarray}
The decay processes under consideration in this paper are summarized in Table \ref{TABLE1}.
Here we can neglect the contribution from resonances which have a mass larger than that of $\Delta(1232)$.
Due to the singularity in the Jacobian $J(p_l,\phi;{\bf V}_{R})$ \cite{HIRANO}, it is difficult to evaluate the momentum distribution in the usual manner. So we use a simple Monte Carlo method to obtain the contribution from the resonance decays.
The basic idea of the method is as follows:

(1) For each fluid element of freeze-out hyper-surface $d\sigma^\mu(x)$, we generate the ensemble of resonance particles in the fluid rest system which obeys the Bose (or Fermi) distribution with freeze-out temperature $T(x)$ and chemical potential $\mu(x)$.

(2) Each resonance particle, which constitutes the thermal distribution, is Lorentz-boosted by the hydrodynamic flow ${\bf v}(x)$.

(3) We make each resonance particle decay under the exact treatment based on the relativistic decay kinematics.

(4) We count the number of decay particles which enter a momentum window under consideration and obtain the spectrum.

\noindent
For details, see also Appendix.

\section{HADRON SPECTRA}\label{HS}
Our strategy to obtain the hadron spectra in non-central collisions is as follows.
First, we choose the initial parameters so as to reproduce the rapidity and transverse mass (momentum) distribution for the most central bin.
Here the transverse profiles of energy density and baryon density are based on the wounded nucleon model. 
Next, we calculate the spectra in non-central collisions by shifting the impact parameter $b$ and without changing the other initial parameters.
Almost all the two-dimensional hydrodynamic calculations are based on the ansatz of wounded nucleon scaling \cite{OLLI92,NUT,KOLB}.
Although the wounded nucleon scaling for the initial condition is widely employed, its validity has not been confirmed yet in terms of the hydrodynamic model. 
In this section, we check whether the ansatz is really valid for the initial condition of the hydrodynamic model by analyzing the spectra in non-central collisions.

\subsection{Central collisions}
\begin{figure}
\begin{center}
\includegraphics[width=12cm]{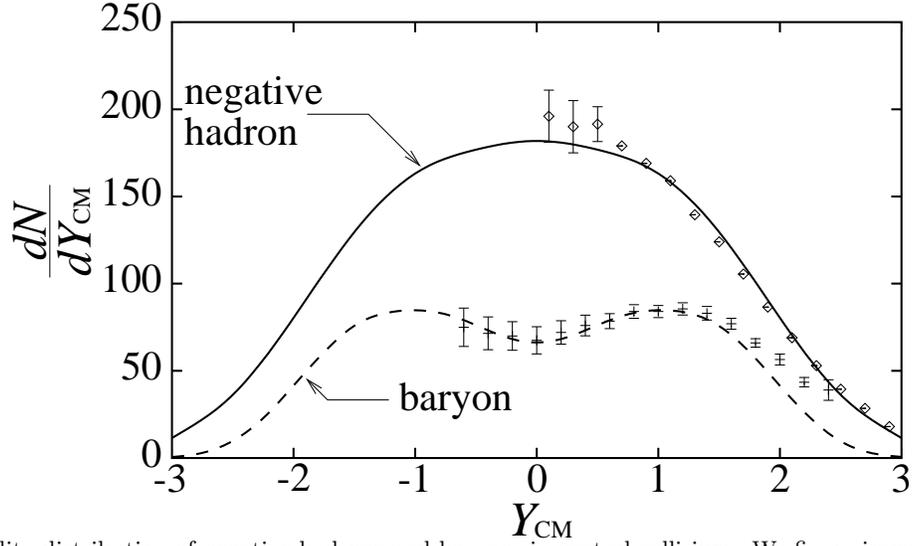}
 \caption{The rapidity distribution of negative hadrons and baryons in central collisions. We fix an impact parameter $b = 2.4$ fm. The solid (dashed) line represents the rapidity distribution of negative hadrons (baryons). }
 \label{RAPCENT}
\end{center}
 \end{figure}

\begin{figure}
\begin{center}
\includegraphics[width=12cm]{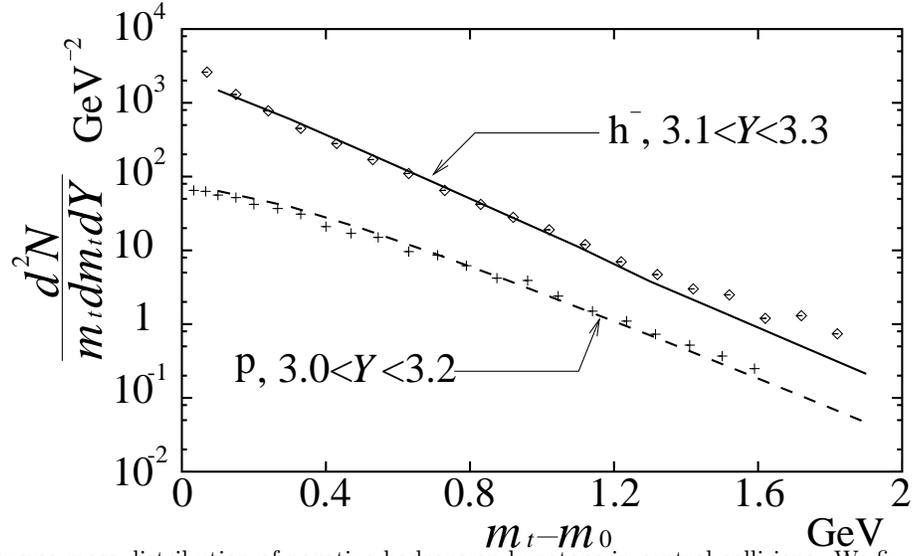}
 \caption{The transverse mass distribution of negative hadrons and protons in central collisions. We fix an impact parameter $b = 2.4$ fm. We adjust the freeze-out energy density $E_{\mathrm{f}}$ to obtain reasonable agreement with the slope of each spectrum. The resultant $E_{\mathrm{f}}$ is 60.0 MeV/fm$^3$.
}
 \label{MTCENT}
\end{center}
 \end{figure}

First, we choose the initial condition so as to reproduce the spectra in central collisions.
We take the four initial parameters in the hydrodynamic simulations as $E_0 = 3.9$ GeV/fm$^3$, $n_{B0} = 0.46$ fm$^{-3}$, $\tilde{z}_0 = 1.38$ fm, and $\alpha = 1.7$. When $\alpha$ is fixed, $n_{B0}$ is automatically determined by Eq.~(\ref{NB0}).
Freeze-out energy density is fixed as $E_{\mathrm{f}} = 60$ MeV/fm$^3$. 
Figure \ref{RAPCENT} represents the numerical results of rapidity distributions for negative hadrons and baryons in Pb + Pb 158 $A$ GeV collisions.
For each distribution, we accumulate the particle directly emitted from the freeze-out hyper-surface and the particles from decay processes represented in Table \ref{TABLE1}.
Experimental data for the 5 \% most central events were obtained by the NA49 Collaboration \cite{NA49RAP}.
By adjusting the initial parameters in the hydrodynamic model, we largely reproduce the experimental data for negative hadrons and net baryons in central collisions.
The multiplicity of anti-protons in our estimate is about unity, so that we safely neglect the contribution of anti-proton within the model calculation.\footnote{The thermal and chemical equilibrium are assumed in the conventional hydrodynamic model. 
On the other hand, it is often said that the chemical freeze-out temperature is larger than the thermal freeze-out temperature in Pb + Pb collisions at the SPS energy.
If we take into account the additional chemical potentials \cite{SHURYAK} for all particles at the temperature between the chemical and the thermal freeze-out, the multiplicity of anti-protons may become larger than the one in the present paper.
See also Ref.~\cite{RAPP}.
} 
In this calculation, we fix the impact parameter $b = 2.4$ fm.
When one analyzes the experimental data in central collisions within a hydrodynamic model, the impact parameter is usually chosen to be zero.
On the other hand, the experimental group always averages over events with small but non-zero impact parameters to obtain the spectra in ``central" collisions.
This is the reason why we choose the impact parameter as 2.4 fm rather than zero.
It should be noted that the 5 \% most central events in Pb + Pb 158 $A$ GeV collisions corresponds to the impact parameter range $0 < b < 3.4$ fm \cite{COOPER}.
The multiplicity of protons within our model EOS is exactly the same as that of neutrons.
On the other hand, the two multiplicities are not equal in Pb + Pb collisions since the real nuclear matter may be isospin asymmetric as mentioned in Sec.~\ref{EOS}.
Hence we analyze the rapidity distribution of net baryons rather than that of net protons.
Figure \ref{MTCENT} represents the transverse mass distribution for negative hadrons and protons.
We found that we obtain a reasonable result for the slope of transverse mass distribution when we choose $E_{\mathrm{f}} = 60$ MeV/fm$^3$.
The corresponding mean thermal freeze-out temperature and chemical potential are, respectively, $<T_{\mathrm{f}}> \sim 117$ MeV and $<\mu_{\mathrm{f}}> \sim 323$ MeV.
The (mean) freeze-out temperature obtained here is consistent with others \cite{NA49RAP,SHURYAK,TF1,TF2}.

\subsection{Non-central collisions}
\begin{figure}
\begin{center}
\includegraphics[width=12cm]{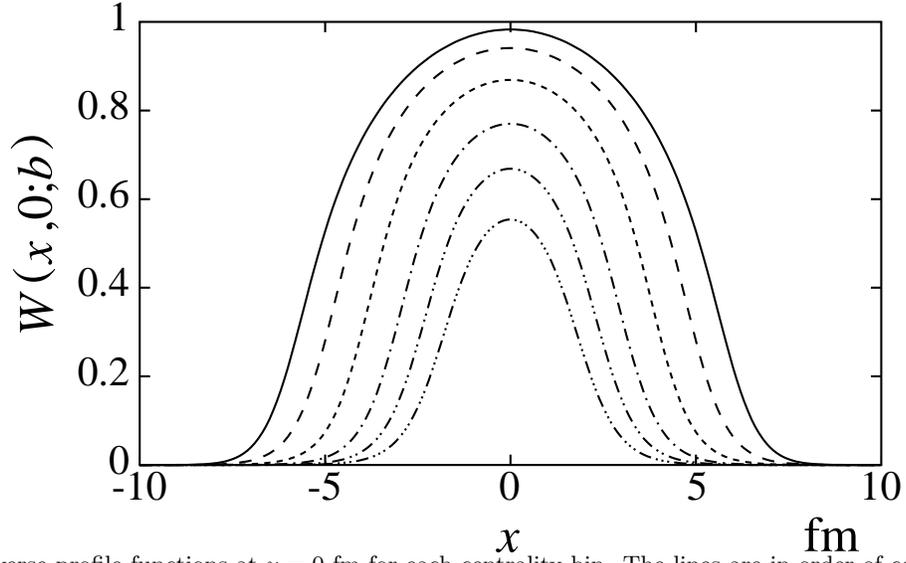}
 \caption{The transverse profile functions at $y = 0$ fm for each centrality bin. The lines are in order of centrality from top to bottom.}
 \label{TRANSPRO}
\end{center}
 \end{figure}

The adjustment of initial condition for central collisions leads us to discuss the non-central collisions through the ansatz of wounded nucleon scaling.
Let us proceed to discussion on non-central events in terms of our hydrodynamic model.
The NA49 Collaboration reported the analysis of non-central events and discussed the centrality dependence of baryon stopping \cite{COOPER}. 
They divided the events into six centrality bins and estimated the impact parameter range for each centrality by using the Glauber model.
When analysing these data, we fix an impact parameter in the hydrodynamic simulation for each centrality bin and the other initial parameters are the same as those in central collisions.
The impact parameter range which is estimated by the experimental group, the fixed impact parameter we choose, the number of wounded nucleons $N_W$, and the maximum values of transverse profile function $W(0, 0; b)$ for each centrality are shown in Table \ref{TABLE2}.
The transverse profile functions at $y = 0$ fm are also represented in Fig.~\ref{TRANSPRO}.
As the impact parameter increases, the maximum value of the transverse profile function decreases as expected.

\begin{figure}
\begin{center}
\includegraphics[width=12cm]{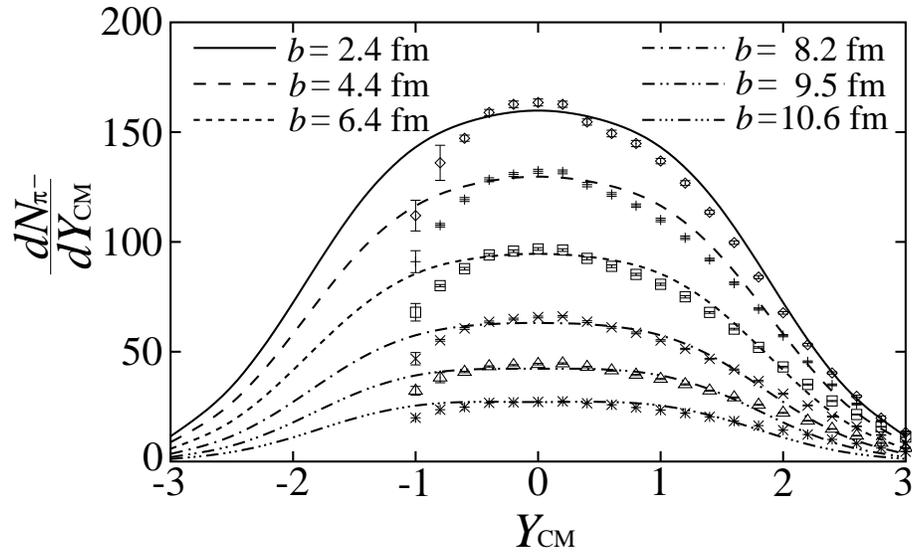}
 \caption{The rapidity distributions of negative pions in non-central collisions for each centrality.
}
 \label{RAPNONCENT}
\end{center}
 \end{figure}

\begin{figure}
\begin{center}
\includegraphics[width=12cm]{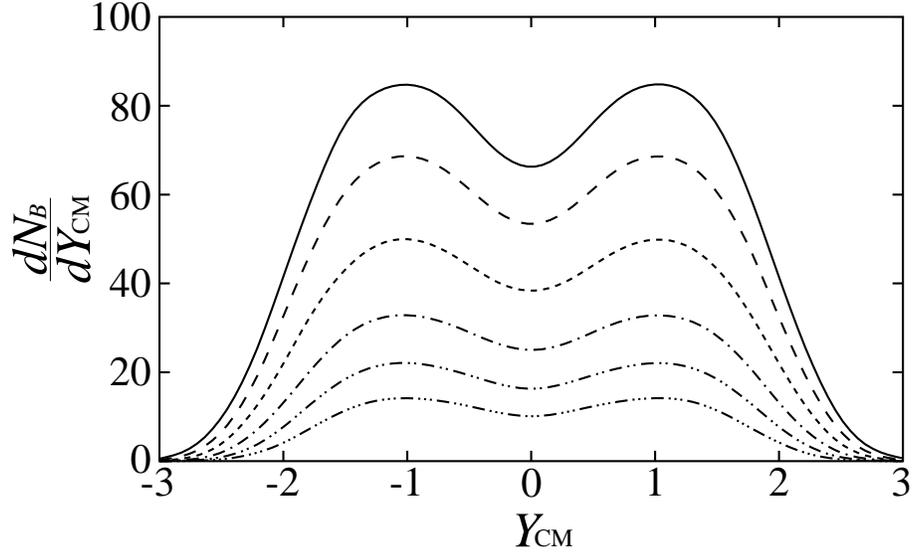}
 \caption{The rapidity distributions of baryons in non-central collisions for each centrality.}
 \label{BRAPNONCENT}
\end{center}
 \end{figure}

\begin{figure}
\begin{center}
\includegraphics[width=12cm]{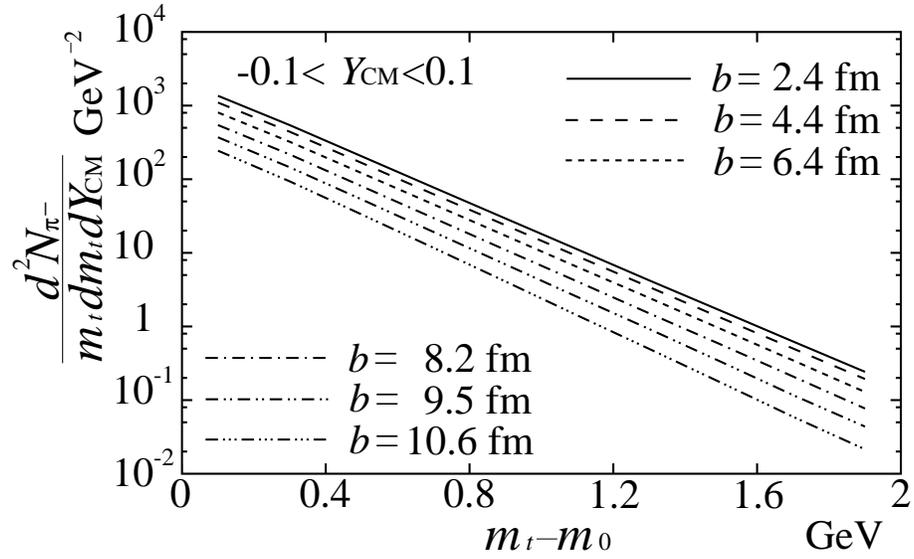}
 \caption{The transverse mass distributions of negative pions in non-central collisions for each centrality.}
 \label{MTNONCENT}
\end{center}
 \end{figure}

\begin{figure}
\begin{center}
\includegraphics[width=12cm]{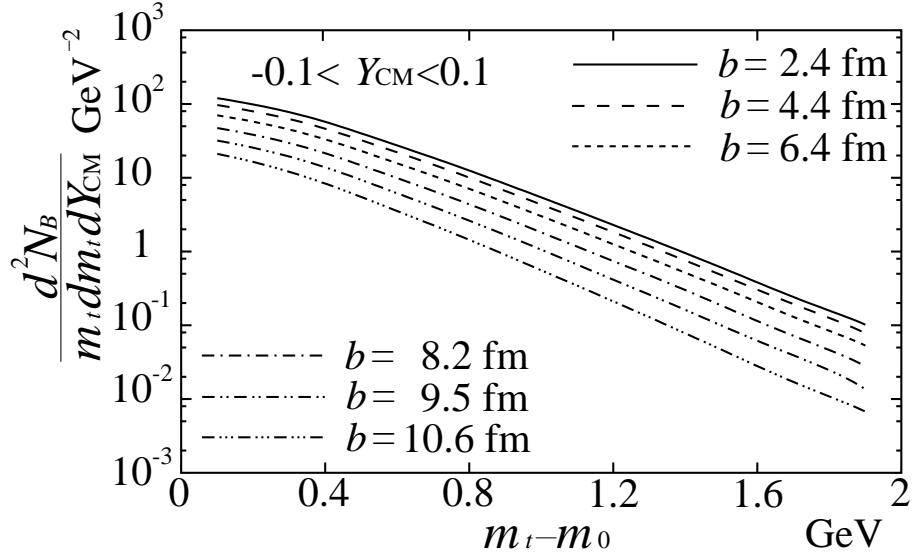}
 \caption{The transverse mass distributions of baryons in non-central collisions for each centrality.}
 \label{BMTNONCENT}
\end{center}
 \end{figure}

We first discuss the pion spectra in non-central collisions.
Figure \ref{RAPNONCENT} represents the rapidity distribution for negative pions in non-central Pb + Pb 158 $A$ GeV collisions.
Our numerical results for each centrality are represented by lines.
These results are in reasonable agreement with the experimental data near midrapidity.
As for the pion spectra, we see from Fig.~\ref{RAPNONCENT} that the ansatz of wounded nucleon scaling works well and seems to be a proper initial condition for non-central collisions in hydrodynamic simulation.
We next discuss the baryon spectra.
Figure~\ref{BRAPNONCENT} shows the rapidity distribution of baryons in non-central collisions. From Fig.~\ref{BRAPNONCENT}, we found that our initialization of baryon density leads to almost constant baryon stopping.
On the other hand, the experimental data shows that the peak position in the rapidity distribution depends on centrality \cite{COOPER}, i.e., baryon stopping increases with centrality.
This implies that, as for the initial \textit{baryon} density, the simple ansatz of wounded nucleon scaling is not enough to describe the impact parameter dependence of baryon stopping.
We also represent in Figs.~\ref{MTNONCENT} and \ref{BMTNONCENT} the results of transverse mass distribution for negative pions and baryons near midrapidity in non-central collisions.  We see the slope of the transverse mass distribution becomes a little steeper as the impact parameter increases. 

\section{SUMMARY AND DISCUSSION}\label{SD}
We analyzed the non-central relativistic heavy-ion collisions by using the (3+1) dimensional hydrodynamic model with a strong first order phase transition between the QGP phase and the hadron phase.
By adjusting the initial parameters in the hydrodynamic model and choosing the impact parameter $b = 2.4$ fm, we reproduced rapidity distributions for negative hadrons and net baryons in central Pb + Pb 158 $A$ GeV collisions. We found that we reproduce the slope of transverse mass distribution with the freeze-out energy density $E_{\mathrm{f}} = 60.0$ MeV/fm$^3$. The corresponding average thermal freeze-out temperature and chemical potential are, respectively, $<T_{\mathrm{f}}> \sim 117$ MeV and $<\mu_{\mathrm{f}}> \sim 323$ MeV. The freeze-out temperature obtained in this paper is consistent with the analyses of the NA49 Collaboration \cite{NA49RAP,TF1}.
We also reproduced the experimental data of rapidity distribution for negative pions in non-central Pb + Pb collisions with the initial condition based on wounded nucleon scaling.
The ansatz of wounded nucleon scaling is reasonable for discussion on the initial condition of energy density at the SPS energy.
On the other hand, we need to modify the initial condition of baryon density in order to explain the centrality dependence of baryon stopping.
We note that the pion spectra are not sensitive to the modification of initial baryon density.
We comment on the initial condition of the longitudinal flow.
Recently, Csernai and Rohrich discussed the `third flow' component of nuclear matter in non-central collisions \cite{CSERNAI}.
In the realistic evolution of nuclear matter, the steepest direction of pressure gradient is tilted from both $x$ and $z$ axes.
This causes anti-flow near the midrapidity region.
In our parametrization, the initial longitudinal flow is independent of transverse coordinates $x$ and $y$ for simplicity.
Thereby we cannot discuss this phenomenon at present.
If we take into account the $x$ dependence of initial longitudinal flow, for example, the initial longitudinal flow increasing near (and decreasing far from) the spectators, we can describe the directed and third flow in our model. 

Reproduction of single particle spectra of hadrons is very important when we try to discuss other topics or make new predictions by using the hydrodynamic model. Therefore we strongly propose that the experimental groups at the SPS and the RHIC should analyze the single particle spectra in \textit{non-central} collisions before they discuss the anisotropic transverse flow.
The numerical results of hydrodynamic simulation obtained in the present paper are our starting point to discuss the elliptic flow and the nutcrack phenomenon at the SPS energy.
One of the authors has already discussed the reduction of elliptic flow due to resonance decays by using the result of the present paper \cite{HIRANO}. We will discuss the centrality dependence of elliptic flow at the SPS energy and how large the reduction effect is in a forthcoming paper \cite{TSUDA}.

\begin{center}
\textbf{ACKNOWLEDGMENTS}
\end{center}

The authors are much indebted to Professor I.~Ohba and Professor H.~Nakazato for their helpful comments and to the members of the high-energy physics group at Waseda University for fruitful discussions.  
They wish to thank M.~Asakawa for a careful reading of the manuscript and C.~Nonaka for giving us a numerical table of EOS.
One of the authors (T.H.) wishes to acknowledge valuable discussions with M.~Asakawa, T.~Hatsuda, P.~Huovinen, P.~F.~Kolb, T.~Matsui, S.~Muroya, C.~Nonaka, and J.-Y.~Ollitrault. T.H. would like to thank A.~M.~Poskanzer for showing him G.~Cooper's Ph.D.~thesis. 
Our numerical calculations were partially performed by NEC SX-4 at Japan Atomic Energy Research Institute (JAERI).
T.H. also thanks S.~Chiba for giving him an opportunity to use NEC SX-4 at JAERI.
T.H.'s research is supported by Waseda University Grant for Special Research Projects No.~2000A-534.

\appendix
\section{MONTE CARLO CALCULATION OF THE CONTRIBUTION FROM RESONANCE DECAYS}
\label{MC}
Lorentz transformation for the momentum of a decay particle between the local rest system (starred) and the finite momentum system (non-starred) of a resonance particle $R$ is 
\begin{eqnarray}
\label{LOTR}
{\bf p}^* = {\bf p}-{\bf p}_{R}\left(\frac{E}{m_{R}}-\frac{{\bf p}\cdot {\bf p}_{R}}{m_{R}(m_{R}+E_{R})} \right).
\end{eqnarray}
We rewrite Eq.~(\ref{LOTR}) explicitly 
\begin{eqnarray}
p^*_l &=& p_l-p_{Rl}F(p_l,\phi), \\
\cos \phi^* &=& \frac{p_x^*}{p_t^*} \nonumber\\
            &=& \frac{p_t(p_l, \phi) \cos \phi-p_{Rt}\cos \phi_{R}F(p_l,\phi)}{\sqrt{p_t^2(p_l, \phi)+p_{Rt}^2 F^2(p_l,\phi)-2p_t(p_l, \phi)p_{Rt}\cos(\phi-\phi_{R})F(p_l,\phi)}},
\end{eqnarray}
where
\begin{eqnarray}
F(p_l,\phi) = \frac{E(p_l, \phi)}{m_{R}}-\frac{p_t(p_l,\phi) p_{Rt}\cos(\phi-\phi_{R})+p_l p_{Rl}}{m_{R}(m_{R}+E_{R})}.
\end{eqnarray}
Here the independent variables which we choose for decay particles are the longitudinal momentum $p_l$ and the azimuthal angle $\phi$.
Thus the transverse momentum of a decay particle $p_t$ is written in terms of $p_l$ and $\phi$:
\begin{eqnarray}
p_t(p_l, \phi) & = & \frac{1}{\gamma_{R}\Bigl(1-v_{Rt}^2\cos^2(\phi-\phi_{R})
\Bigl)}
\biggl((E^* + p_l v_{Rl} \gamma_{R})v_{Rt}\cos(\phi-\phi_{R})\nonumber \\
& \pm & \sqrt{(E^* + p_l v_{Rl} \gamma_{R})^2-(p_l^2+m^2)\gamma_{R}^2(1-v_{Rt}^2\cos^2(\phi-\phi_{R}))} \biggl).
\end{eqnarray}

The Jacobian of the Lorentz transformation is defined by
\begin{eqnarray}
dp_l^*d\phi^* & = & J(p_l,\phi;{\bf V}_{R})dp_ld\phi,\\
\label{JACOBI}
J(p_l,\phi;{\bf V}_{R}) & = & \left|
 \begin{array}{cc}
  \frac{\partial p_l^*}{\partial p_l} & \frac{\partial p_l^*}{\partial \phi}\\
  \frac{\partial \phi^*}{\partial p_l} & \frac{\partial \phi^*}{\partial \phi}
 \end{array}
\right|.
\end{eqnarray}
The calculation of $J$ is straightforward, so that we do not represent it here.
The normalization of momentum space volume for a decay particle in the resonance rest frame is 
\begin{eqnarray}
\int_{-p^*}^{p^*}\frac{dp_l^*}{2p^*} \int_0^{2 \pi} \frac{d\phi^*}{2 \pi} = 1.
\end{eqnarray}
We always average the decay probability over the spin of resonances, so that the decay probability does not depend on $p_l^*$ and $\phi^*$.
Thus the normalization in the resonance reference frame is
\begin{eqnarray}
\int\frac{J(p_l,\phi;{\bf V}_{R})dp_ld\phi}{4\pi p^*} = 1. 
\end{eqnarray}

The Jacobian in Eq.~(\ref{JACOBI}) has very narrow peaks when the resonance particle moves at a large velocity in the laboratory system \cite{HIRANO}. Due to this singularity, it is very difficult to integrate the Jacobian numerically. So we introduce a very simple Monte Carlo calculation to evaluate the momentum distribution from resonance decays.
All input parameters in this calculation are the numerical results of hydrodynamic simulation, the temperature $T$, the chemical potential $\mu$, the three-dimensional fluid velocity ${\bf v}$, and the element of freeze-out hyper-surface $d \sigma^\mu$ on the freeze-out hyper-surface $\Sigma$.
In the following discussion, we show how to obtain the rapidity distribution of negative pions, for simplicity, only from $\rho$-mesons.
In this case, the branching ratio $B_{\rho^{0(-)} \rightarrow \pi^- \pi^{+(0)}} = 1$.
It is straightforward to extend this scheme to the cases for other resonances or the transverse mass (momentum) distribution.

\noindent
\textit{Step 1}: Evaluate the number of $\rho^0$ and $\rho^-$ which are emitted from \textit{or} absorbed by the $k$-th freeze-out hyper-surface element $d \sigma^\mu_k$:
\begin{eqnarray}
N^{R}_k = \frac{g_{R}}{(2 \pi)^3}\int \frac{d^3p_{R}}{E_{R}} \frac{\mid p_{R\mu} d\sigma^\mu_k \mid}{\exp(p_{R}^\nu u_\nu^k/T_k) - 1}.
\end{eqnarray}
The integrand does not contain the Jacobian, so that it is simple to carry out the numerical integration by a standard technique.
It should be noted that $N^{R}_k$ is different from the \textit{net} number of emitted $\rho$-mesons from the $k$-th fluid element.

\noindent
\textit{Step 2}: Generate $\tilde{N}$ random momentums $P^*_j$ ($1\le j \le \tilde{N}$) for $\rho$-mesons which obey the distribution
\begin{eqnarray}
\frac{{P^*}^2}{\exp(\sqrt{{P^*}^2+m^2_{R}}/T_k) - 1}.
\end{eqnarray} 
Here, for simplicity, we omit the Breit-Wigner function.

\noindent
\textit{Step 3}: For each $\tilde{N}$ random momentum $P^*_j$, generate random variables $(\Theta^*_j, \Phi^*_j)$ whose ensemble is uniformly distributed on the unit sphere.
By using these random variables, we obtain an ensemble of $\rho$-meson with momentum ${\bf P}^*_j = (P^*_{xj}, P^*_{yj}, P^*_{zj})= (P^*_j \sin\Theta^*_j \cos \Phi^*_j, P^*_j \sin\Theta^*_j \sin \Phi^*_j, P^*_j \cos \Theta^*_j)$, which obeys the Bose-Einstein distribution in the fluid rest system.

\noindent
\textit{Step 4}: Boost ${\bf P}^*_j$ with respect to the fluid velocity ${\bf v}_k$:
\begin{eqnarray}
{\bf P}_j = {\bf P}^*_j+{\bf v}_k \gamma_k \left(E^*_j+\frac{{\bf P}^*_j\cdot {\bf v}_k\gamma_k}{1+\gamma_k} \right),
\end{eqnarray}
where $\gamma_k = 1/\sqrt{1-{\bf v}_k^2}$.

\noindent
\textit{Step 5}: Generate $\tilde{N}$ uniform random variables on the unit sphere  $(\theta^*_j, \phi^*_j)$ and obtain an ensemble of pion with momentum ${\bf p}^*_j = (p^*_{xj}, p^*_{yj}, p^*_{zj}) = (p^* \sin\theta^*_j \cos \phi^*_j, p^* \sin\theta^*_j \sin \phi^*_j, p^* \cos \theta^*_j)$, where $p^*$ is given by Eq.~(\ref{PSTAR}).

\noindent
\textit{Step 6}: Boost ${\bf p}^*_j$ with respect to the resonance momentum ${\bf P}_j$:
\begin{eqnarray}
{\bf p}_j = {\bf p}^*_j+{\bf P}_j\left(\frac{E_j}{m_{R}}+\frac{{\bf p}^*_j \cdot {\bf P}_j}{m_{R}(m_{R}+E_{R})} \right).
\end{eqnarray}

\noindent
\textit{Step 7}: If $P^\mu_j d\sigma_{\mu k}$ is positive, 
\begin{eqnarray}
N^+_k \rightarrow N^+_k + \frac{P^\mu_j d\sigma_{\mu k}}{E_j^*},
\end{eqnarray}
if $P^\mu_j d\sigma_{\mu k}$ is negative,
\begin{eqnarray}
N^-_k \rightarrow N^-_k + \frac{\mid P^\mu_j d\sigma_{\mu k} \mid}{E_j^*}.
\end{eqnarray}
Here, $N^+_k$ ($N^-_k$) is to be proportional to the number of $\rho$-mesons which are emitted from (absorbed by) the $k$-th fluid element.

\noindent
\textit{Step 8}: If the rapidity of a negative pion $Y_j$ which is evaluated from ${\bf p}_j$ enters in a rapidity window $Y-\frac{\Delta Y}{2} < Y_j < Y+\frac{\Delta Y}{2}$ and $P^\mu_j d\sigma_{\mu k}$ is positive, then
\begin{eqnarray}
\Delta N^+_k(Y) \rightarrow \Delta N^+_k (Y)+\frac{P^\mu_j d\sigma_{\mu k}}{E_j^*},
\end{eqnarray}
if $Y_j$ also enters the above rapidity window but $P^\mu_j d\sigma_{\mu k}$ is negative,
\begin{eqnarray}
\Delta N^-_k(Y) \rightarrow \Delta N^-_k(Y)+\frac{\mid P^\mu_j d\sigma_{\mu k} \mid}{E_j^*}.
\end{eqnarray}

\noindent
\textit{Step 9}: Repeat steps 7 and 8 for all $\tilde{N}$ random variables.

\noindent
\textit{Step 10}: Obtain the rapidity distribution of decay particles from the $k$-th
fluid element
\begin{eqnarray}
\label{DNKDY}
\frac{dN_k}{dY}(Y) = \frac{N^R_k}{N^+_k + N^-_k}\left(\frac{\Delta N^+_k(Y) - \Delta N^-_k(Y)}{\Delta Y} \right).
\end{eqnarray}
It should be noted that the minus sign in the bracket of Eq. (\ref{DNKDY}) means the \textit{net} number of emitted particles from the $k$-th fluid element.
For the normalization in Eq. (\ref{DNKDY}), we use the gross number $N_k^+ + N_k^-$ since this number is positive definite.

\noindent
\textit{Step 11}: Repeat the above steps from 1 to 10 for all fluid elements obtained in a numerical simulation of the hydrodynamic model.
Summing over the contribution from all fluid elements on the freeze-out hyper-surface $\Sigma$, we obtain the rapidity distribution of $\pi^-$ which are from $\rho$ decays:
\begin{eqnarray}
\frac{dN_{\rho \rightarrow \pi^- X}}{dY}(Y) = \sum_k \frac{dN_k}{dY}(Y).
\end{eqnarray}


%
%
 
%
%
\begin{table}
\caption{Resonance decays which contribute to the spectrum. The fractions in the branching ratio come from the Clebsch-Gordan coefficients.}
\label{TABLE1}
\begin{tabular}{llc}
Decay particle & Decay channel & Branching ratio\\
\hline
$\pi^-$ & $\rho^0 \rightarrow \pi^- \pi^+$ & $1.0$ \\
& $\rho^- \rightarrow \pi^- \pi^0$ & $1.0$ \\
& $\omega \rightarrow \pi^- \pi^0 \pi^+$ & $0.88$ \\
& $K^{*0} \rightarrow \pi^- K^+$ & $\frac{2}{3} \times 1.0$ \\
& $K^{*-} \rightarrow \pi^- \bar{K}^0$ & $\frac{2}{3} \times 1.0$ \\
& $\Delta^- \rightarrow \pi^- n$ & $1.0$ \\
& $\Delta^0 \rightarrow \pi^- p$ & $\frac{1}{3} \times 1.0$ \\
$K^-$ & $K^{*-} \rightarrow K^- \pi^+$ & $\frac{1}{3} \times 1.0$ \\
& $\bar{K}^{*0} \rightarrow K^- \pi^0$ & $\frac{2}{3} \times 1.0$ \\
$p$ & $\Delta^0 \rightarrow p \pi^-$ & $\frac{1}{3} \times 1.0$ \\
& $\Delta^+ \rightarrow p \pi^0$ & $\frac{2}{3} \times 1.0$ \\
& $\Delta^{++} \rightarrow p \pi^+$ & $1.0$ \\
$n$ & $\Delta^{+} \rightarrow n \pi^+$ & $\frac{1}{3} \times 1.0$ \\
& $\Delta^{0} \rightarrow n \pi^0$ & $\frac{2}{3} \times 1.0$ \\
& $\Delta^{-} \rightarrow n \pi^-$ & $1.0$ \\
\end{tabular}
\end{table}

 \begin{table}
 \caption{The impact parameters in the hydrodynamic simulations. $N_W$ represents the number of wounded nucleons. $W$ is a transverse profile function. For details, see text.}
 \label{TABLE2}
 \begin{tabular}{ccccccc}
& Bin1 & Bin2 & Bin3 & Bin4 & Bin5 & Bin6\\
\hline
range (fm) & 0-3.4 & 3.4-5.4 & 5.4-7.4 & 7.4-9.1 & 9.1-10.2 & 10.2- \\
$b$ (fm) & 2.4 & 4.4 & 6.4 & 8.2 & 9.5 & 10.6 \\
$N_W$ & 353 & 282 & 199 & 128 & 83 & 51 \\
$W(0, 0; b)$ & 0.983 & 0.941 & 0.869 & 0.770 & 0.669 & 0.554 \\
 \end{tabular}
 \end{table}

\end{document}